\documentclass[aps,prd,nofootinbib,twocolumn,floatfix,groupedaddress]{revtex4}
\usepackage{amsfonts,amssymb,epsfig,verbatim,mathbbol,amstext,amsmath,psfrag}

\def\tr{\mbox{tr}}

\def\eq{Eqn.}
\def\eqs{Eqns.}
\def\fig{Fig.}

\def\f{f}

\def\j{l}

\begin{document}

\title{Dressed Wilson loops as dual condensates in response to magnetic and electric fields}

\author{Falk Bruckmann}

\author{Gergely Endr\H{o}di}
\affiliation{Institut f\"ur Theoretische Physik, Universit\"at Regensburg, D-93040 Regensburg, Germany.}

\begin{abstract}
We introduce dressed Wilson loops as a novel confinement observable. It consists of closed planar loops of arbitrary geometry but fixed area, and its expectation values decay with the latter. The construction of dressed Wilson loops is based on chiral condensates in response to magnetic and electric fields, thus linking different physical concepts. We present results for generalized condensates and dressed Wilson loops on dynamical lattice configurations and confirm the agreement with conventional Wilson loops in the limit of large probe mass. We comment on the renormalization of dressed Wilson loops.
\end{abstract}
 

\maketitle

\section{Introduction}

For the analysis of confining gauge theories Wilson and Polyakov loops are the key ingredients. Quark confinement reveals itself in an area law for the former and a va\-ni\-shing expectation value for the latter. This is because these loops represent non-Abelian factors along worldlines of heavy quarks. Consequently, the logarithms of the Wilson and Polyakov loop are related to the interquark potential and the free energy of a single quark at finite temperature, respectively.

In a previous work, dressed Polyakov loops were introduced \cite{Bilgici:2008}, which opened a new perspective on the relation between confinement and chiral symmetry. Dressed Polyakov loops are center sensitive like conventional Polyakov loops and hence order parameters of the deconfinement transition \cite{Bilgici:2010_Zhang:2010}. Moreover, they connect to chiral symmetry since they are defined as Fourier components of quark condensates with respect to phase boundary conditions in the temporal direction \cite{Gattringer:2006}, reflected in the name `dual condensate'.

Dressed Polyakov loops contain a tunable probe mass $m$, which allows us to explore a number of their interesting features. In the limit of large mass one can expand the observable in inverse powers of the mass in terms of closed loops. Long loops are suppressed by the mass, which therefore prefers straight loops and makes contact with the conventional Polyakov loop in this limit. In turn, a finite mass governs the distance over which `detours' in the dressed Pol\-ya\-kov loop extend. The probe mass also controls which part of the Dirac spectrum dominates the condensate. Furthermore, dressed Polyakov loops receive less renormalization than conventional ones \cite{Bruckmann:2008b}, cf.\ \cite{Synatschke:2008}.

In this work we introduce dressed Wilson loops defined as (collections of) planar closed loops of fixed area, but variable geometry. We view these loops as enclosing the same amount of disorder as conventional Wilson loops, which should reflect itself in an area law. In the same manner as for dressed Polyakov loops, dressed Wilson loops are defined using a tunable probe mass parameter. Large values of the probe mass can be shown to suppress long loops preferring `ideal' loops, which on the lattice we demonstrate to be rectangular and thus make contact to conventional Wilson loops.
 
Dressed Wilson loops are constructed via the quark condensate in the background of external abelian gauge fields. In this way, space-time and spatial Wilson loops are connected to the response of the quark condensate to (Euclidean) electric and magnetic fields, respectively, which 
gives an intuitive physical picture to be explored further. 
Similar magnetic fields play an important role at current heavy ion colliders and are investigated in relation to the QCD phase diagram. (Constant magnetic fields can be used to realize noncommutative coordinates, too, Wilson loops in this setting were calculated in \cite{Bietenholz:2007}.)

The definition of dressed Wilson loops through electric and magnetic fields should also  help to access these nonlocal objects in diagrammatic approaches and QCD models, as is the case for dressed Polyakov loops \cite{Fischer:2009g_Braun:2009_Kashiwa:2009}.

This paper is organized as follows. First we introduce external fields and define generalized and dual condensates/dressed Wilson loops, both in the continuum and on the lattice. After that, in Sect.~\ref{sect lattice}, we present our numerical results from lattice simulations. Sect.~\ref{sect further} is devoted to further conceptual remarks on our observable. In Sect.~\ref{sect geometry} we discuss the geometry of lattice loops, which enables us to make contact to conventional Wilson loops in Sect.~\ref{sect contact}. We conclude with remarks on the renormalization in Sect.~\ref{sect renorm} and an outlook.

\section{Dressed Wilson loops: idea, definition}

The concept of dressed Wilson loops is based on the possibility to incorporate the area of any (planar) loop in factors of the links along the loop. This is achieved by an auxiliary field strength, which is constant and abelian, through Stokes' theorem.

To consider planar loops in two fixed directions $\mu$ and $\nu$, let $a_\mu$ and  $a_\nu$ denote abelian gauge fields generating a constant field strength $\f$ with these indices,
\begin{equation}
f_{\mu\nu}=\partial_\mu a_\nu-\partial_\nu a_\mu\equiv \f \stackrel{(!)}{=}\mbox{constant}
\quad \mbox{for given } \mu,\nu 
\end{equation} 
A possible choice on the infinite plane is $a_\nu \!=\! \f x_\mu,a_\mu\!=\!0 $.

For every given nonabelian configuration from the QCD path integral we modify the original nonabelian fields (indicated by capital letters) by
\begin{equation}
A_\mu\to A_\mu+a_\mu\,,\quad
 A_\nu\to A_\nu+a_\nu\,,\quad
F_{\mu\nu}\to F_{\mu\nu}+\f
\label{eqn mod nonab}
\end{equation}
(in these directions only). Consequently, a closed loop on a contour $C$ bounding an area $S$ is multiplied by 
\begin{equation}
 e^{i\oint_C a_\mu dx_\mu}
=e^{i\int\!\!\!\int_S f d^2\sigma}
=e^{i \f S}
\label{eqn Wilson factors}
\end{equation}
which commutes through the nonabelian part (we set the charge to unity). The area $S$ is oriented (from the contour $C$), inverse loops are multiplied by the inverse factor. 
These modified links are to be used in a physical quantity, e.g.\ the condensate, and its dual quantity after Fourier transform in $\f$ will consist of Wilson loops of fixed area $S$. 

In order to specify the details, let us consider from now on the finite volume setting, i.e.\ the two directions being periodic with extensions $L_\mu$ and $L_\nu$ spanning a total area $S_{\mu\nu}\equiv L_\mu L_\nu$. Then the flux of abelian fields -- like momentum -- obeys a quantization condition \cite{tHooft:1979uj},
\begin{equation}
\f=2\pi\,\frac{k}{S_{\mu \nu}}\equiv f_k\,,
 \qquad k\in\mathbb{Z}\,,
\label{eqn f quant}
\end{equation} 
in terms of the area, see e.g.\ \cite{Al-Hashimi:2009}.

We start with the generalized quark condensate 
\begin{equation}
\Sigma_k\equiv \frac{1}{S_{\mu\nu}} \, \big\langle \tr\, \frac{1}{D_k+ m}\big\rangle
\label{eqn sigma k}
\end{equation}
where $D_k$ is the Dirac operator evaluated with the external abelian field of strength $\f_k$ from \eq~(\ref{eqn f quant}) and restricted to the $(\mu,\nu)$-plane. The trace stands for a sum over color and spin degrees of freedom plus an integral over space normalized by the two-volume $S_{\mu\nu}$ (such that $\tr$ is dimensionless and later on the lattice it simply amounts to a sum over all sites). Note that like for the four-dimensional condensate $\Sigma$ is an intensive quantity, but with dimension of mass (not mass cubed). 

From gauge invariance it is clear that this observable consists of closed loops. Furthermore, the trace ensures that only connected loops contribute to $\Sigma_k$ (the loops are on the other hand allowed to self-intersect). Of these loops, we are interested only in non-winding loops (and only for those loops Stokes' theorem applies). We therefore assume to have control over the contribution from winding loops, e.g.\ by a Fourier transform wrt. boundary conditions as for the dressed Polyakov loop (now in both compact directions). The role of winding loops will be elaborated on in more detail later.

As argued above, closed loops in the generalized quark condensate obtain factors according to their area $S\in[0,S_{\mu\nu}]$ in the $(\mu,\nu)$-plane and $\Sigma_k$ is decomposed as
\begin{equation}
\Sigma_k=\int_0^{S_{\mu\nu}}\!\!dS\: \tilde{\Sigma}(S)\,e^{i\f_kS}
\label{eqn decomp cont} 
\end{equation}
This is nothing but a Fourier transform with the abelian fields $\f_k$ and the area $S$ as conjugate variables, which is in complete analogy to boundary condition angle and winding number for the dressed Polyakov loop. 

We invert this formula to obtain the \emph{dual condensate}
\begin{equation}
 \tilde{\Sigma}(S)\equiv\frac{1}{S_{\mu\nu}}\,\sum_{k\in \mathbb{Z}}e^{-i\f_k S}\Sigma_k
 \label{eqn sigma s}
\end{equation}
which consists of loops having fixed area $S$. This defines the \emph{dressed Wilson loop}. We constructed it to have dimension mass cubed. The geometries of the contained loops are arbitrary, they contribute to $\tilde{\Sigma}(S)$ with weights depending on $m$ and the underlying space, see below.

To calculate this novel observable on a discrete lattice, we have to specify the definitions a bit further. With lattice spacing $a$ and $N_{\mu,\nu}=L_{\mu,\nu}/a$ points in the two directions, respectively, the area of any loop is of course quantized as $S=sa^2$ with integer $s\in[0,N_{\mu\nu}\equiv N_\mu N_\nu]$. The magnetic flux is also bounded from above,
\begin{equation}
k_{{\rm max}}=\frac{S_{\mu \nu}}{a^2}=N_{\mu \nu}\,, 
\end{equation} 
(the value at which each plaquette receives a factor $1$), such that all lattice quantities are $N_{\mu\nu}$-periodic in $k$. The external abelian fields can be implemented on the lattice by multiplying the link variables with appropriate space-dependent complex phases, see e.g. \cite{D'Elia:2011}.

\begin{figure}[t]
\includegraphics[width=\linewidth]{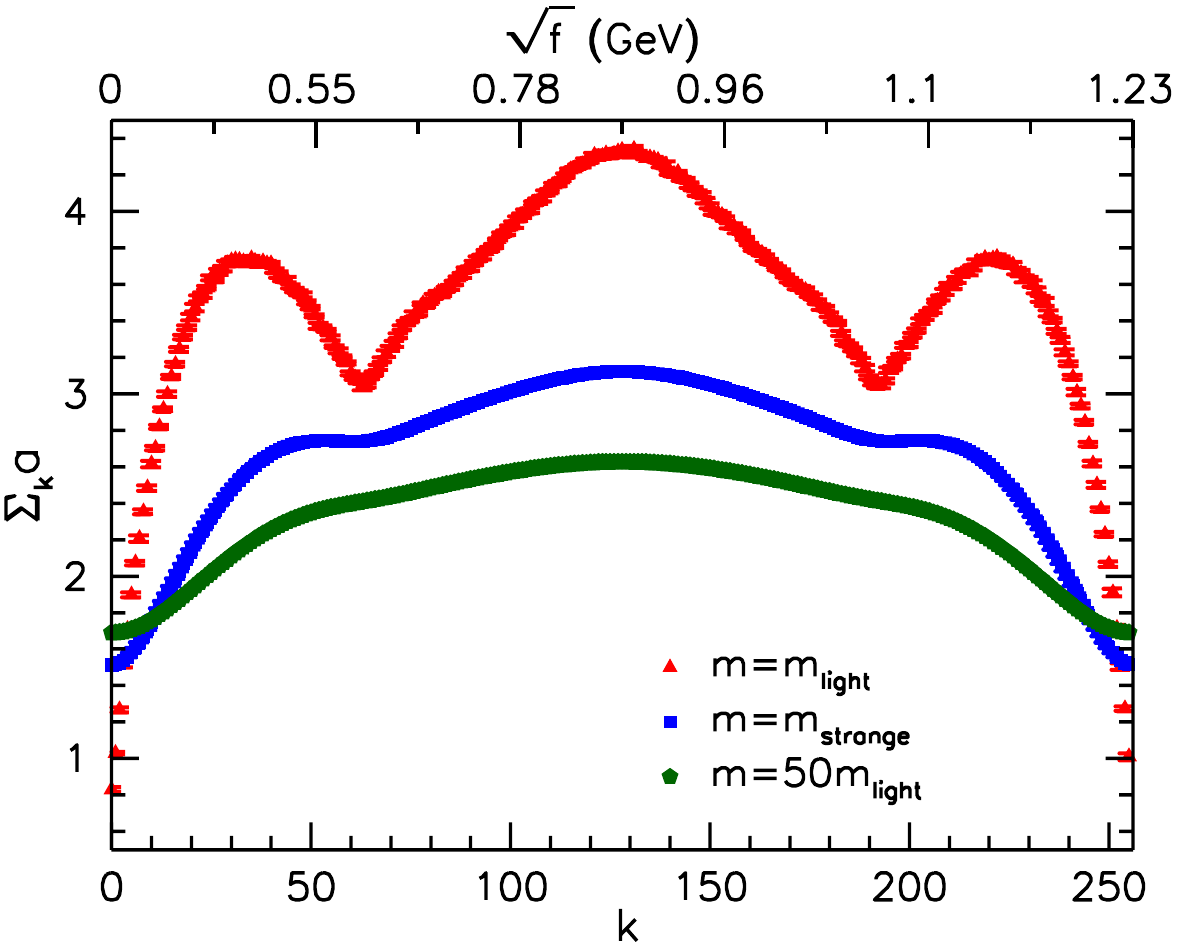} 
\caption{The generalized quark condensate, \eq~(\protect\ref{eqn sigma k}), in response to a constant magnetic field as a function of the flux quantum for several probe masses. On the upper axis we also show the magnetic field in physical units.}
\label{fig sigma k}
\end{figure}

The abelian factor for a Wilson loop of area $s$ is $\exp(i f_k S)=\exp(2\pi i ks/N_{\mu\nu})$,
where the flux and area quanta $k$ and $s$ are dual variables running over the same discrete range. The dual condensate/dressed Wilson loop is again defined via a Fourier transform, now the discrete one
\begin{equation}
 \tilde{\Sigma}(s)\equiv\frac{1}{N_{\mu\nu}a^2}\,\sum_{k=1}^{N_{\mu\nu}}
 e^{-2\pi iks/N_{\mu\nu}}\Sigma_k
 \label{eqn sigma s lattice}
 \end{equation}
(such that $\tilde{\Sigma}(0)$ is the average of $\Sigma_k$'s divided by $a^2$).

\section{Lattice results}
\label{sect lattice}

For our lattice studies we used the Symanzik improved gauge action and $N_f=2+1$ flavors of stout smeared staggered fermions with physical masses. We generated 5 configurations (which due to self-averaging effects of the Wilson loops sufficed for our purposes) on $16^3\times 4$ lattices at 
$\beta=3.31$ ($a\approx 0.4$~fm, $T\approx 120$~MeV). Further details of the simulation setup can be found in e.g.~\cite{Aoki:2006br}. 
The condensates, \eq~(\ref{eqn sigma k}), were calculated from the spectra of $D_k$, cf.~\eq~(\ref{eqn spec rep}); we computed all its eigenvalues at all possible magnetic fields with LAPACK.

\begin{figure}[t!]
\includegraphics[width=1.03\linewidth]{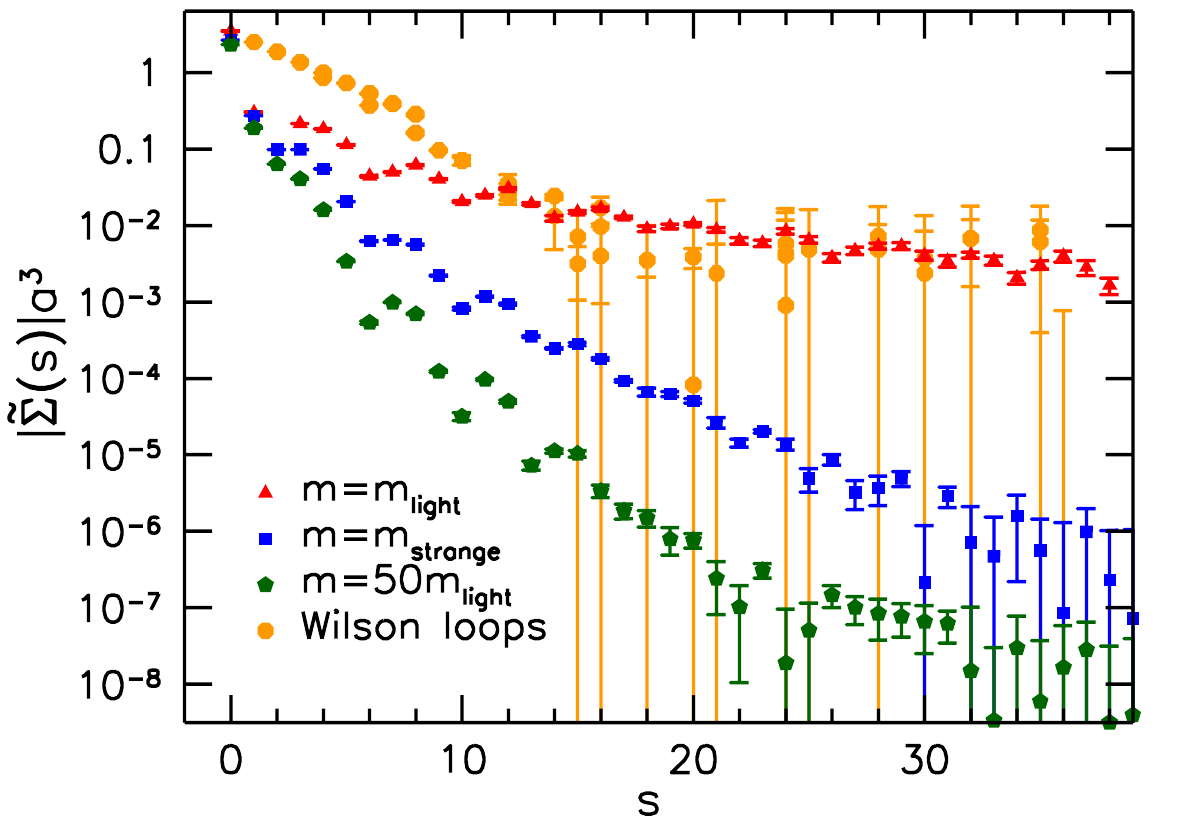} 
\caption{The (absolute value of) dressed Wilson loops, \eq~(\protect\ref{eqn sigma s lattice}), dual to the condensates of \fig~\protect\ref{fig sigma k} together with conventional Wilson loops  (the latter measured for different rectangles with side lengths up to $N_s/2$).}
\label{fig sigma s}
\end{figure}

We present results for the staggered Dirac operator
and the case of a constant magnetic field in the $z$ direction, i.e.\ for Wilson loops in the $(x,y)$-plane. The generalized quark condensate $\Sigma_k$ as a function of the magnetic field is shown in Fig.~\ref{fig sigma k}. The condensate at $k=0$, the conventional one without magnetic field,  increases with the mass (for small masses) as is well-known. For nonzero $k$ we expect that heavy particles are less affected by external fields and indeed the $\Sigma_k$ curve flattens out. The qualitative behavior of $\Sigma_k$ is similar if one uses the 4d Dirac operator in the definition (not shown, for this check we used random estimators to determine the condensate). 

The dressed Wilson loops $\tilde{\Sigma}$ corresponding to the condensates shown in Fig.~\ref{fig sigma k} are plotted as a function of area in Fig.~\ref{fig sigma s}. They reveal the expected decay in area similar to conventional Wilson loops. Our observable contains all kinds of loop geometries and therefore has an enhanced signal/noise ratio  compared to conventional Wilson loops (by up to an order of magnitude for low masses). The decay rate as well as some modulations in its dependence on area, however, renders it different from conventional Wilson loops, to which we will come back below.

The presented data are based on all closed loops, i.e.\ also winding loops. Since the number of the latter is only proportional to the circumference of the lattice (while the number of all the loops is proportional to the area) we expect winding loops to give negligible contributions to $\tilde\Sigma$ for small probe masses. To confirm this we repeated the measurements under all combinations of periodic and antiperiodic boundary conditions in the two directions. Averaging these results (which can be considered the crudest approximation to the zeroth Fourier component) excludes loops with odd winding numbers. We observe that the results are within statistical
errors unchanged.

\section{Further conceptual remarks}
\label{sect further}

The applied constant abelian field $\f_{\mu\nu}$ is sensitive to loops in the $(\mu,\nu)$-plane. This means that electric and magnetic fields correspond to space-time and spatial Wilson loops, respectively. While at zero temperature the QCD vacuum is 4d-symmetric, at finite temperature a difference is expected in the loops and correspondingly in the response to those fields.

The abelian field is not sensitive to the `4d area' of loops extending in directions other than $(\mu,\nu)$. Such detours would largely spoil the identification of loops of same area. Therefore, since we want keep contact to conventional planar Wilson loops of some area, we have restricted all generalized and dual observables, in particular the Dirac operator, to some fixed  $(\mu,\nu)$-plane. The calculation of its spectrum on the lattice is thus very cheap. This, however, does not mean that the system reduces to a 2d gauge theory, since the gauge fields in the observable have interacted via the full action and hence embody 4d dynamics. Naturally, the calculation in a fixed plane is followed by an average over all such planes, just as for conventional Wilson loops. 

We emphasize that for the computation of dressed Wilson loops all possible abelian fields need to be integrated over in the Fourier transform. (Typical magnetic fields of phenomenological relevance for heavy ion colliders are physically very large, but correspond in current lattice simulations to the smallest possible quanta $k=O(1)-O(10)$.) We also stress, that our setting is partially quenched in the sense that observables are measured on configurations without external field in the fermionic action (which should explain some of the differences to measurements in e.g.\ \cite{D'Elia:2011}).

An intuitive physical picture of our construction can be drawn from 
Wilson loops extending in space and time (say in the $(x,t)$-plane), which represent a quark-antiquark pair. The corresponding dressed Wilson loops are obtained from an external electric field ($E_x$) instead of a magnetic one discussed so far. 
In the common picture of confinement the quark and the antiquark are bound together by nonabelian forces, that -- if one was able to separate them to a  large distance $R$ -- result in a potential $V(R)$ growing linearly with $R$. Our (Euclidean) electric field acts on the quark and antiquark, as they carry opposite electric charges, and thus indeed attempts to vary the distance $R$ between them. Therefore it is clear that the electric field probes the nonabelian forces. Our formalism says that the latter are exhibited from the response of QCD quantities to all such electric fields (similar to inverse scattering).

The QCD vacuum is symmetric under inverting the external field $\f\to-\f$, i.e. $k\to -k$. The finite lattice resolution and the associated flux periodicity lead to the symmetry $\Sigma_k=\Sigma_{N_{\mu\nu}-k}$, obeyed on average, see \fig~\ref{fig sigma k}. In order to obtain dressed Wilson loops at large areas, the high frequency fluctuations in $\Sigma_k$ need to be computed. Their calculation is limited by the flux quantization in finite volumes. The finite volume also limits large conventional Wilson loops.

In the spectral representation of the Dirac operator $D_k$ with eigenvalues $i\lambda_k^{(i)}$, that are purely imaginary and come in complex conjugate pairs (neglecting nongeneric zero modes),
\begin{equation}
 \tr\, \frac{1}{D_k+ m}=\sum_i\frac{1}{i\lambda_k^{(i)}+m}=
\sum_{i, \lambda_k^{(i)}>0} \frac{2m}{\lambda_k^{(i)\,2}+m^2}
\label{eqn spec rep}
\end{equation}
it is obvious that the dominant contribution to all condensates comes from the IR part $\lambda\lesssim m$ (for the analogous discussion in dressed Polyakov loops see \cite{Bilgici:2010_Zhang:2010}). 

An important effect in dressed Wilson loops is the suppression of long loops by heavy probes. For large $m$ one can expand the condensate into a geometric series,
\begin{equation}
 \tr\, \frac{1}{D_k+ m}=\frac{1}{m}\sum_{\j=0}^\infty (-1)^\j\frac{\tr\, D_k^\j}{m^\j}
 \label{eqn geom series}
\end{equation}
For a lattice Dirac operator hopping over just nearest neighbors like the staggered operator we use, $\j$ is the length of the loop in lattice units, the length in physical units being $L=la$. To generate a closed loop, $\j$ needs to be even and the sign on the right hand side of \eq~(\protect\ref{eqn geom series}) can be removed. 

From \eq~(\ref{eqn geom series}) it is clear that loops will be suppressed by the associated factor of inverse $m$ (the same argument would apply to the scalar propagator from the lattice Laplacian). Thus we expect dressed Wilson loops to be `fuzzy' with the inverse mass governing their width.
We will demonstrate this now on the lattice.

\section{Geometry of lattice loops} 
\label{sect geometry}

In the following we specialize to the staggered Dirac operator again. The signs in this operator yield signs in the dressed Wilson loops depending on their length and area as follows. A closed loop consists of an even number $\j$ of links, of which $\j/2$ are inverse links $U^\dagger$ each coming with a minus sign. For the total staggered phase one can tile the area of the loop into plaquettes and convince oneself that odd areas get an additional minus sign, altogehter giving rise to a sign factor of 
$(-1)^{s+l/2}$.
In order not to complicate the notation, in the equations $\tilde\Sigma$ denotes the absolute value of the dual condensate (this also applies to the dual
power of the Dirac operator $\tilde{T}$, defined later).

What is the `ideal lattice loop' in the sense of maximizing the area at given circumference? For every concave loop the hooks can be folded to the outside keeping or even lowering the circumference,\\

\begin{center}
\includegraphics[height=0.12\linewidth]{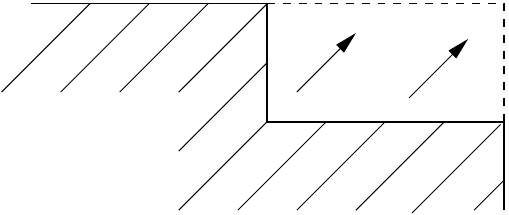}\qquad
\includegraphics[height=0.12\linewidth]{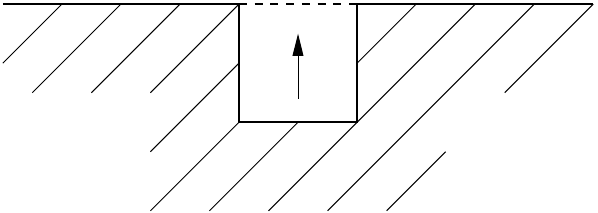}
\end{center}
and the limits of this procedure are \emph{rectangles} (among rectangles with same circumference those closest to squares have maximal area), contrary to the continuum, where the ideal loops are circles.

The `entropy', the multiplicity of different loops in the observable, may counteract the suppression away from the ideal loop. To check for this effect we define a combinatorial factor
\begin{equation}
 F(s,\j)\equiv\frac{\#\mbox{ loops of area }s\mbox{ and circumference }\j}{N_{\mu\nu}} 
\label{eqn def F}
\end{equation}
Note that the loops can start at any point in the lattice plane, therefore we divide by the lattice area to arrive at an intensive quantity. Furthermore, the circumference is meant in a general sense as the number of lattice bonds needed to traverse a loop of area $s$ (since the link can, e.g., go back and forth). 

We calculated $F$ using a recursive algorithm. This combinatorial factor apparently arises in different contexts \cite{Bhattacharya:1997,Mashkevich:2009}. At low area we reproduced the results of \cite{Mashkevich:2009} and extended them up to higher circumferences/areas, see the appendix. At fixed area and large circumference $F$ grows exponentially,
\begin{equation}
 F(s,\j)\approx \frac{4^\j}{\j^2} \quad\mbox{ for }\j \gg s
\label{eqn F large length}
\end{equation}
The increase in this multiplicity factor (at fixed area $s$) is always exceeded by the increase in the suppressing mass factor $m^\j$ for large enough mass (note that the color factor from the loop depends on $\j$, too, but cannot grow exponentially). Thus we conclude that the dual condensate $\tilde{\Sigma}(s)$ at large mass contains predominantly loops of small circumference.

\section{Contact to conventional Wilson loops}
\label{sect contact}
 
From the considerations in the last sections the dressed Wilson loops in the large mass limit should be proportional to conventional Wilson loops. Note that in this limit the UV eigenvalues now dominate, see also \cite{Bruckmann:2007b}.

From \eq~(\ref{eqn geom series}) it is clear that the leading contribution in $\tilde{\Sigma}(s)$ comes from the Dirac operator raised to the minimal circumference for the given area, denoted by $\j_{{\rm min}}(s)$, in other words, $\j_{{\rm min}}(s)$ is the first number $\j$ of link hoppings that is able to generate an area $s$ (see also the appendix). Loops with areas $s$ and circumference $\j_{{\rm min}}(s)$ describe certain geometries. For square number areas $s=p^2$ these are square loops $p\times p$. The starting point can be anywhere on the square and therefore the combinatorial factor $F$ is the number of points on the square 
\begin{equation}
 F(p^2,4p)=4p
\label{eqn F square}
\end{equation}
For areas $s=p(p+1)$ the minimal loops are rectangles $p\times (p+1)$, where the longer side can point in both directions on the plane. Accordingly, the combinatorial factor is twice the number of points
\begin{equation}
 F(p(p+1),4p+2)=8p+4
\label{eqn F rectangle}
\end{equation}
In other cases the geometry of loops with minimal circumference is not unique, but always `close to rectangular'.

For powers of the Dirac operator we introduce the shorthand notation,
\begin{equation}
 T^{\j}_k\equiv\frac{1}{S_{\mu\nu}}\big\langle\tr\,D^{\j}_k\big\rangle
\end{equation}
which is an intensive quantity, and its Fourier transform $\tilde{T}^l(s)$ is obtained like in  \eq~(\ref{eqn decomp cont}). Then for large probe masses
\begin{equation}
 m\,\tilde{\Sigma}(s)\rightarrow
\frac{\tilde{T}^{\j_{{\rm min}}(s)}(s)}{m^{\j_{{\rm min}}(s)}}
+O\Big(\frac{\tilde{T}^{\j_{{\rm min}}(s)+2}(s)}{m^{\j_{{\rm min}}(s)+2}}\Big)
 \label{eqn SDm}
\end{equation}
where the correction term comes from the loop with same area but circumference increased by 2. This correction enters both the power of the Dirac operator and the mass suppression factor. 

For Dirac operators containing only $U/2a$ and $U^{(\dagger)}/2a$ one has the following relation to Wilson loops
\begin{equation}
 \tilde{T}^{\j_{{\rm min}}(s)}(s)=
\frac{1}{a^4}
\frac{F(s,\j_{{\rm min}}(s))}{(2a)^{\j_{{\rm min}}(s)}}N_c
\big\langle W(s)\big\rangle
 \label{eqn DFW}
\end{equation}
The color factor $N_c$ comes about in order to take into account the $1/N_c$ prefactor in the usual definition of traced Wilson loops $W$.

Specializing to square loops, $s=p^2$ with $\j_{{\rm min}}=4p$ and $F$ from \eq~(\ref{eqn F square}) we obtain
\begin{equation}
\tilde{T}^{4p}(p^2)=
\frac{1}{a^4}
\frac{4p}{(2a)^{4p}}N_c
\big\langle W(p\times p)\big\rangle
 \label{eqn DFW square}
\end{equation} 
Combining that with \eq~(\ref{eqn SDm}) and reverting to physical circumference (`length') of Wilson loops  $L=4pa$ and side length $R=pa=L/4$, conventional square Wilson loops can be recovered from
\begin{equation}
 m\,\tilde{\Sigma}(R^2)\rightarrow
\frac{1}{a^4}
\frac{L/a}{(2am)^{L/a}}N_c
\big\langle W(R\times R)\big\rangle\Big(1+O\Big(\frac{1}{(2am)^2}\Big)\Big)
\label{eqn SFW}
\end{equation}
where the first factor on the right hand side accounts for the dimension of the dual condensate and the last one indicates the first correction from non-ideal loops being suppressed in powers of $2am$ (with coefficient that depends on the length).

\begin{figure}[t]
\includegraphics[width=\linewidth]{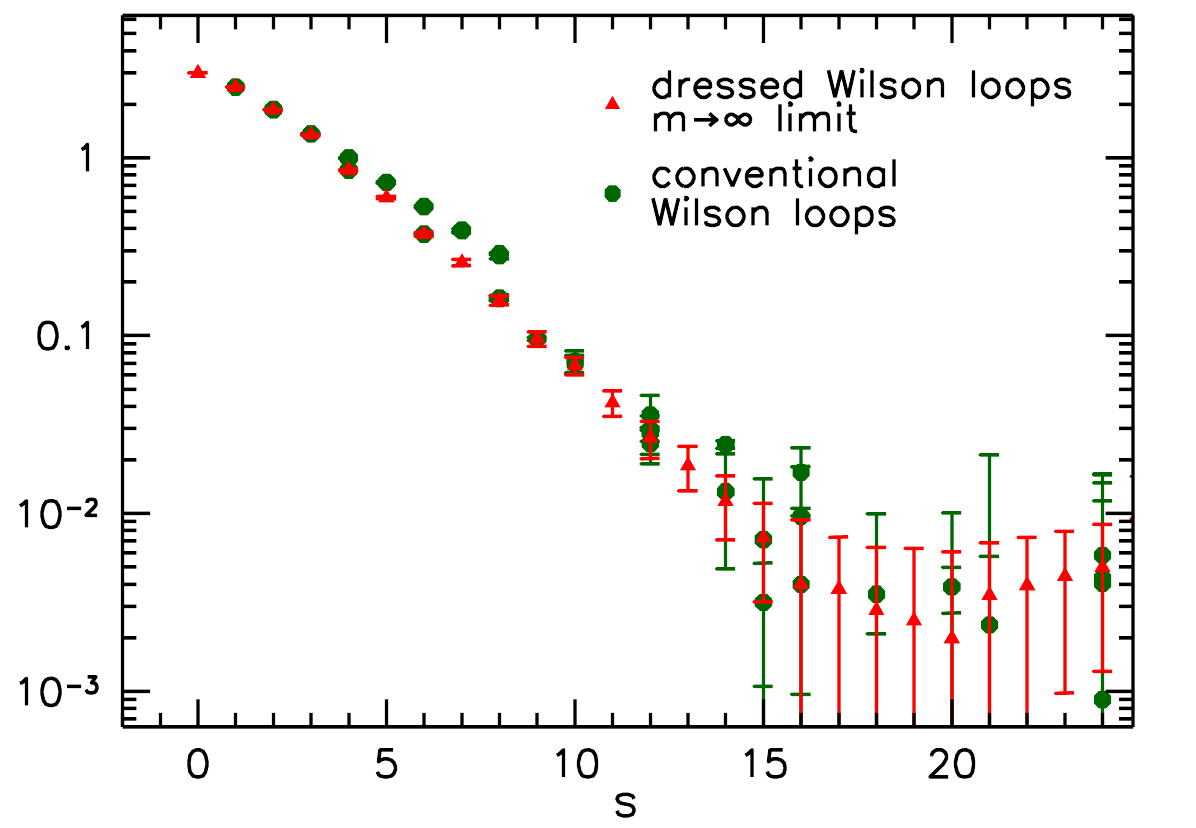}
\caption{The leading term in the large mass limit of the dressed Wilson loop, (absolute value of) $\tilde{T}^{\j_{{\rm min}}(s)}(s)$, rescaled by the according combinatorial factor $F$ and the other factors in \eq~(\protect\ref{eqn DFW}), in comparison to conventional Wilson loops.}
\label{fig sigma s largemass}
\end{figure}

In order to prove the equivalence of the large mass limit of
dressed loops and conventional loops numerically, $\tilde{T}^{\j_{{\rm min}}(s)}$ from \eq~(\ref{eqn DFW}) is better suited than $m\tilde{\Sigma}$ because the latter is tiny due to the according mass factor, see \eq~(\ref{eqn SDm}). We show in \fig~\ref{fig sigma s largemass} the numerical results for $\tilde{T}^{\j_{{\rm min}}(s)}$ from \eq~(\ref{eqn DFW}) divided by the accompanying factors (note that the mass has dropped out of this relation) and see that indeed the conventional Wilson loops are reproduced. We note that one can prove that in this limit the contamination from winding loops becomes critical at $s=(N_s/4)^2+N_s/4+1$ in the sense that winding loops contributing to this $s$ can be shorter than $\j_{\rm min}(s)$. By filtering out loops with odd winding number once again, this critical area can be pushed up beyond $N_s^2/4$, where conventional Wilson loops are also out of reach.

\section{Remarks on Renormalization}
\label{sect renorm}

From the connection of dual condensates to conventional Wilson loops at fixed lattice spacing it is tempting to obtain the renormalization of the latter\footnote{The renormalization of Wilson loops is usually done via particles moving on them \cite{Polyakov:1980ca_Dotsenko:1979wb} and is sensitive to edges of the Wilson loop.} from the former, too. 

First of all, ordinary condensates defined with the 4d Dirac operator are known to be subject to additive and multiplicative renormalization. Dressed Wilson loops consist of subsets of loops present in 4d condensates, but feel the same lattice spacing and therefore should contain the same UV-divergences.
For dual condensates one then expects that additive renormalization is not necessary since these divergences do not depend on the external field\footnote{We remind the reader that in our approach the external field only enters on the level of observables.} and therefore are removed by the Fourier transform (as long as $s\neq 0$). The multiplicative renormalization can easily be done by multiplying the condensate by the bare mass $m$, which has already been done in the previous section. 

Under the conjecture that $m\,\tilde{\Sigma}(s)$ is a renormalized quantity, \eq~(\ref{eqn SFW}) seems to reveal the $a$-dependent factors needed to renormalize conventional Wilson loops\footnote{and indeed the length dependence of these terms is of the form obtained in \cite{Polyakov:1980ca_Dotsenko:1979wb}.}. We remind the reader that this connection to Wilson loops is valid only for large $am$, which is of course related to the use of unphysical infinitely heavy probe quarks in conventional Wilson loops. In the continuum limit of condensates at fixed physical mass, however, the bare mass needs to run such that the combination $am$ goes to zero \cite{Montvay:1994cy}. 
More work is needed to understand this situation (e.g.\ whether it also appears in other regulators than lattice) and the role of dressed Wilson loops.

\section{Conclusions and outlook}
 
We have presented the definition and first results for dressed Wilson loops. They indeed obey a confining area law and in the limit of large probe mass reproduce conventional Wilson loops. To demonstrate this we have used lattice simulations, but we strongly expect our observable to be useful in continuum approaches as well.

Being constructed from the response of quark condensates to magnetic and electric fields, dressed Wilson loops should help to get more insight into the interrelation of these concepts. This concerns in particular the behavior at the QCD transition. Further interesting aspects of dressed Wilson loops, partly deduced from the analogy to dressed Polyakov loops, are their IR dominance and renormalization properties. 
Moreover, the possibility to use dressed Wilson loops for setting the scale is attractive.%

\section*{Acknowledgements}

The authors thank the Budapest-Wuppertal collaboration for the permission to use their staggered code for the configuration production, and G.~Bali, C.~Gattringer, M.~G\"ockeler and A.~Sch\"afer for helpful discussions. FB likes to thank the organizers of the `Workshop on Strongly-Interacting Field Theories' in Jena, where the idea for this work was born. This work has been supported by DFG (BR 2872/4-2) and the Research Executive Agency 
of the European Union 
(ITN STRONGnet).

\newpage

\appendix

\begin{widetext} 

\section*{Appendix: Combinatorial factors}

\renewcommand{\tabcolsep}{0.12cm}

In the following table we list the combinatorial factor $F(s,l)$, the number of possible loops with $l$ two-dimensional lattice hoppings (`circumference', but hoppings can be forth and back) that span an area of $s$ lattice units, divided by the total number of lattice points, cf.~\eq~(\ref{eqn def F}), up to $l=24$. Note that the area is oriented and thus can be negative, the combinatorial factor is the same, $F(-s,l)=F(s,l)$.

\begin{tabular}{r|rrrrrrrrrrrr}
 $s\backslash l$& 2 & 4 & 6 & 8 & 10 & 12 & 14 & 16 & 18 & 20 & 22 & 24\\
\hline
0 & 4 &
28 &
232 &
2156 &
21944 &
240280 &
2787320 &
33820044 &
424925872 &
5486681368 &
72398776344 &
972270849512 \\
1 && {\bf 4} &
72 &
1008 &
13160 &
168780 &
2168544 &
28133728 &
369612648 &
4920045800 &
66324542240 &
904584355488 \\
2 &&& {\bf 12} &
308 &
5540 &
87192 &
1291220 &
18569808 &
263462220 &
3718483560 &
52450578224 &
741275064780 \\
3 &&&& 48 &
1560 &
33628 &
610232 &
10127744 &
159762240 &
2445203460 &
36746804160 &
546205048128 \\
4 &&&& {\bf 8} &
420 &
11964 &
262612 &
5015108 &
88145244 &
1469700900 &
23667392012 &
372270358704 \\
5 &&&&& 80 &
3636 &
101976 &
2289760 &
45306288 &
827935484 &
14348361544 &
239687474784 \\
6 &&&&& {\bf 20} &
1200 &
40376 &
1036368 &
22761228 &
452521560 &
8408087996 &
148916557340 \\
7 &&&&&& 264 &
13720 &
435040 &
10920456 &
239017700 &
4784523128 &
90018016224 \\
8 &&&&&& 72 &
4900 &
184104 &
5208372 &
124781340 &
2683439616 &
53545392516 \\
9 &&&&&& {\bf 12} &
1512 &
73056 &
2398752 &
63687440 &
1479475008 &
31384157616 \\
10 &&&&&&& 420 &
28064 &
1085724 &
32101920 &
807111536 &
18215426820 \\
11 &&&&&&& 112 &
10336 &
478008 &
15887160 &
434604632 &
10465152864 \\
12 &&&&&&& {\bf 28} &
3760 &
208728 &
7803420 &
232384020 &
5973330680 \\
13 &&&&&&&& 1088 &
84312 &
3699300 &
121904288 &
3369519456 \\
14 &&&&&&&& 352 &
34560 &
1749200 &
63575468 &
1889445720 \\
15 &&&&&&&& 96 &
13392 &
805280 &
32649760 &
1048976880 \\
16 &&&&&&&& {\bf 16} &
4788 &
359760 &
16509284 &
576770520\\
17 &&&&&&&&& 1584 &
155560 &
8195968 &
313483392 \\
18 &&&&&&&&& 540 &
66960 &
4037528 &
169191804 \\
19 &&&&&&&&& 144 &
26920 &
1931952 &
89921136 \\
20 &&&&&&&&& {\bf 36} &
10680 &
914540 &
47383128 \\
21 &&&&&&&&&& 3740 &
415888 &
24503088 \\
22 &&&&&&&&&& 1360 &
188144 &
12567024 \\
23 &&&&&&&&&& 440 &
82016 &
6330912 \\
24 &&&&&&&&&& 120 &
34496 &
3138264 \\
25 &&&&&&&&&& {\bf 20} &
13640 &
1517328 \\
26 &&&&&&&&&&& 5236 &
723264 \\ 
27 &&&&&&&&&&& 1936 &
337104 \\
28 &&&&&&&&&&& 660 &
153360 \\
29 &&&&&&&&&&& 176 &
66528 \\
30 &&&&&&&&&&& {\bf 44} &
28464 \\
31 &&&&&&&&&&&& 11280 \\
32 &&&&&&&&&&&& 4488 \\
33 &&&&&&&&&&&& 1632 \\
34 &&&&&&&&&&&& 528 \\
35 &&&&&&&&&&&& 144 \\
36 &&&&&&&&&&&& {\bf 24} 
\end{tabular}
The bold numbers are those with $s=p^2$ or $s=p(p+1)$ and $\j=\j_{{\rm min}}(s)$ used in \eqs~(\ref{eqn F square}) and (\ref{eqn F rectangle}). The geometries of the associated loops are ideal in that they maximize the area at given circumference (therefore in the table these numbers are the lower ends of the $\j$-columns), namely by squares and rectangles, respectively. 

\eq~(\ref{eqn F large length}) on the other hand describes the exponential growth of $F$ in the rows at fixed area $s$. 
\end{widetext}

\end{document}